# ROTATIONAL PERIODS OF MAGNETIC CHEMICALLY PECULIAR STARS

**Zdenek MIKULASEK[1,3], Gabriel SZASZ[1], Jiri KRTICKA[1], Juraj ZVERKO[2], Jozef ZIZNOVSKY[2], Miloslav ZEJDA[1], Tomas GRAF[3]**

[1]*Department of Theoretical Physics and Astrophysics, Masaryk University, Kotlářská 2, Brno, Czech Republic*
[2]*Astronomical Institute, Slovak Academy of Sciences, Tatranská Lomnica, Slovak Republic*
[3]*Observatory and Planetarium of J. Palisa, VSB - Technical University, Ostrava, Czech Republic*

**Abstract**

Magnetic chemically peculiar (mCP) stars of the upper part of the main sequence usually exhibit strictly periodic variations of their light, spectrum and magnetic field. These changes can be well explained using the oblique rotator model of the rigidly rotating star with persistent photometric spots, uneven horizontal abundance of chemical elements and dipole-like magnetic field. Long-term observations of mCP stars variability enable to determine their rotational periods with an extraordinary accuracy and reliability.

We compare rotational periods of mCP stars with those of normal main sequence stars of the same spectral type and discuss the found discrepancies. We also discuss the cases of mCP stars V901 Ori, SX Ari, CU Vir, (and possibly also EE Dra), which have been reported to display an increase of their rotational periods.

## 1. Introduction

Stars are isolated bodies made of high-temperature plasma which is held together by its own gravity. They are formed by collapse of dense cores of interstellar clouds, consisting mostly of hydrogen, along with helium and trace amounts of heavier elements. Stars spend most of their active lives on the main sequence (MS). Those located at the central part of the MS can produce energy in their cores by processes of nuclear fusion, converting hydrogen into helium. This process gradually changes chemical composition of their interiors; however, this change has (in most cases) no influence on their outer layers, where the observed spectra are formed. Therefore, the chemical composition of stellar atmosphere reflects the composition of primordial nebula. Every star also carries part of the nebula's angular momentum and that is the reason why all stars have to rotate.

First main sequence star with confirmed rotation was our Sun. Galileo Galilei, who was systematically sketching sunspot patterns, already interpreted those daily changes on solar surface as consequence of the Sun's rotation. The sunspots are not optimum reference points for determination of the rotational period. They appear, evolve and disappear on time-scales comparable with rotational period of the Sun. Nevertheless, it was soon found out, that the Sun rotates in about 25 days relative to the stars.

In case of the Sun we have an advantage, because we can observe individual spots. That is impossible in the case of distant stars. In theory, it should be possible to determine rotational period from slight photometric variability of the star as whole, but it can be done only in exceptional cases when the stellar activity is very high.
However, global information about rotational periods of single MS stars and its dependency on effective temperature can be obtained spectroscopically. Rotation is the primary cause of the profile broadening of most of the lines in stellar spectra. By their analysis it is possible to determine the $V \sin i$ value, where $i$ is the inclination of rotation axis and $V$ is the equatorial rotational velocity of the star. This velocity depends on the rotational period $P$ (in days) and the stellar radius $R$ (in solar radii) according to the well-known formula:

$$V = 50.613 \,\text{km s}^{-1} \frac{R}{P}.$$

Thus, those parameters can be basically obtained from the spectral line broadening of the rotating stars. Even if it is possible to make a reliable estimate of the MS star radius, the period itself cannot be determined, only its maximal value $P/\sin i$. Assuming that directions of rotation axes of the Galactic stars are distributed randomly, it is possible to derive mean rotational periods of the MS stars with different effective temperatures from the mean $<V \sin i>$ values.

As shown on the figure below, the shortest period (approx. 0.55 day) is typical for stars of spectral types in range B2 V – F0 V. Rotational periods grows up towards both early and late type stars; however, the Sun is abnormally slow rotating star anyway.

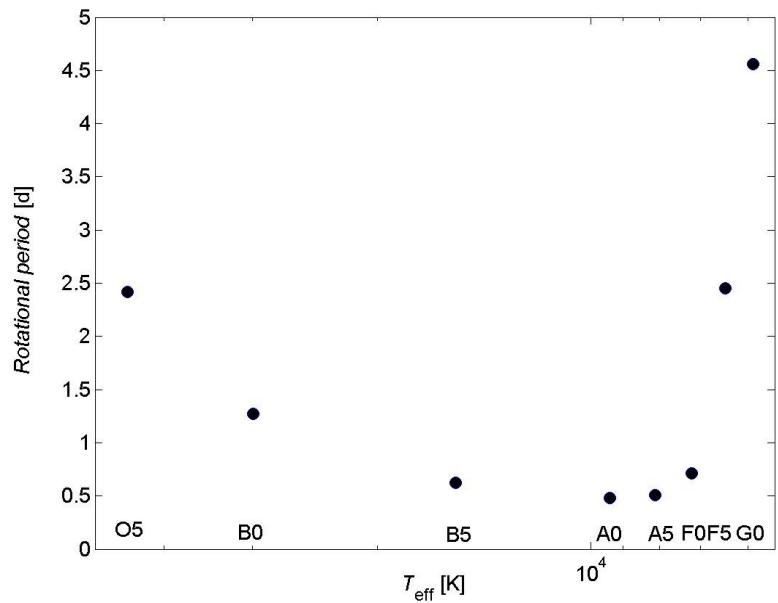

**Figure 1.** Dependency of mean rotational period for main sequence stars on effective temperature/spectral type. $<V \sin i>$ and radii of MS stars were taken from McNally (1965) and Harmanec (1988), respectively.

It was found out that the stellar activity of solar-type stars is usually higher than the activity of the Sun. Its lower activity is most likely caused by abnormally slow rotation. Spots on the surface of late type stars are usually larger and their lifetimes use to be significantly longer than their rotational period. Therefore, there is a large fraction of stars showing photometric variability modulated by rotational period that is caused by the occurrence of surface spots. The rotational periods determined in this way are typically only a couple of days. This result fully agrees with the conclusions based on the analysis of the $<V \sin i>$ values.

Rotation of the MS star that is a component of a close binary system is bound, thus its rotational period is equal to the orbital period of the system. Of course, there are some exceptions, e.g. in case of intensive transfer of mass and angular momentum between the components or in the case of fast change of the stellar interior when the star is approaching to the terminal age main sequence. This is possibly the case of the primary component of Am eclipsing binary HR 6611 (Mikulasek et al. 2004).

## 1.1 Cool and hot stars

Detailed observations of sunspots in different heliographic latitudes have showed up that the Sun does not rotate as a solid body. Angular velocity of the equatorial parts is higher than that in the middle heliographic latitudes or in the polar regions. It seems that phenomenon of differential rotation is symptomatic for all bodies with ongoing convection heat transfer in outer layers. This applies to solar-type stars, as well as red dwarfs, brown dwarfs and giant planets. The first such star, other than Sun, that has its differential rotation mapped in detail is AB Doradus (Donati & Collier Cameron 1997).

Presence of extensive convective layer under surface affects observed behaviour of MS star in crucial way. We can define two physically different groups called hot and cool stars. The division line, not exactly sharp, lies at the effective temperature approx. 7200 K. Photospheres of the cool stars are rather turbulent due to the presence of rising convective currents. Huge magnetic fields, generated by dynamo mechanism, rise towards stellar surface and form here so-called active regions. Dissipation of these magnetic fields then causes various types of stellar activity, i.e. solar-type photospheric spots, flares, prominences, coronal holes.

In contrary, subsurface layers of the hot MS stars, in which heat is transferred by mostly calm radiative diffusion, are utterly stable. Of course, upper layers of the stellar atmospheres with effective temperature higher than about 22 000 K are slightly influenced by the outflow of accelerating stellar winds.

Everything supports the view that hot stars rotate as solid bodies. Their magnetic fields, if any, should be fossil ones and they should survive for the whole existence of these MS stars, according to the higher conductivity of the outer layers and shorter lifetime. Any solar-type stellar activity, including presence of large photospheric spots can be no way expected. Chemical composition should be identical with that of primordial nebula from which the star was formed.

## 2. Chemically peculiar stars

Conclusions above and theoretical expectations perfectly fit the most of the observed hot MS stars; nevertheless, about 10-15% of them show some anomalies. Those are the so-called A and B class peculiar stars (Ap and Bp stars), which significantly differ from the normal stars with the same effective temperature, especially due to their peculiar spectra. These spectra indicate abnormal chemical composition of the atmospheres; therefore, those stars are *chemically peculiar*. Observed chemical anomalies are not related only to rare earths and metals, but also to helium, the second most common element in the universe. They are probably a consequence of slow selective microscopic processes of radiative diffusion and gravitational settling, that proceed in particularly calm outer layers of stars, mostly stabilized by global magnetic field.

Many of the chemically peculiar stars manifest strictly periodic changes in their spectra, effective magnetic induction and brightness. All these changes were satisfactorily explained by the oblique rotator model (ORM) of the CP star (Stibbs 1950) with a dipole-like non-axisymmetric magnetic field, uneven horizontal distribution of chemical elements, and photometric spots in its photosphere. Available observing data suggest that those surface structures are persistent, at least on time scale of decades.

Strict periodicity of the observed light changes of CP stars allows determination of their rotational period with unprecedented accuracy, what gives us a unique possibility to study the rotational characteristics of the hot MS stars and their possible delicate changes. Development of rotational periods would suggest something about evolutionary changes of the stellar bodies and also about interaction of the stars with their neighbourhood. However, it was found out during the second half of the last century, that the CP stars differ from the normal ones by slower rotation as follows: $<P^{-1}>_{norm} \sim 4.8 <P^{-1}>_{CP}$ (as we derived from *On-line catalogue of photometric observation of mCP stars* – see bellow).

It is generally accepted that the slow rotation is the necessary condition (but not the sufficient one) to maintain surface chemical anomaly that would be normally immediately destroyed by meridional currents induced by flattening of the star due to its rotation. Therefore the CP anomaly preferentially develops on the surface of the stars that are slow rotators since their formation. It is also possible that in the case of stars with strong magnetic field their slow rotation by magnetic braking, early after reaching the zero age main sequence. The chemical anomaly then appears right after the rotational period is long enough to allow its formation. However, those are only speculations, and they have to be corroborated by reliable observing evidence. Consequently, the extensive study of the rotational period stability applying on the largest possible sample of CP stars is highly demanding.

## 2.1 On-line catalogue of photometric observations of mCP stars and its exploitation

Nowadays, we know thousands of CP stars in our Galaxy and the other thousands were found in neighbour galaxies, especially in the Large and Small Magellanic Clouds. Only a small fraction of them have known type of their peculiarity and less then one tenth is being observed often enough to determine their rotational period or period of photometric variability. That is mostly determined by precise photoelectric photometry with accuracy better than 0.005 mag[1], due to the typical amplitudes of mCP star light variations (several hundredths of magnitude). The most important source of the high quality photometric data in astrophysically well-defined Strömgren *uvby* system is Saul J. Adelman, who observes mCP stars for decades using FCAPT, the automatic observing facility in Arizona.

Photometric data of the mCP stars are published in different formats in various journals. Lack of a uniform access to the published data makes any systematic study of the photometric behaviour of the mCP stars rather difficult, if not impossible. For that reason we initiated an extensive long-term project in 2004. The goal is to collect all photometric data available to Galactic mCP stars and import them into a single database. The *On-line catalogue of photometric observations of mCP star*s (Mikulasek et al. 2007) is already fully operational for several years and it is publicly accessible via address: http://astro.physics.muni.cz/mcpod/
Nowadays, the archive contains 147 705 individual measurements of 151 mCP stars and it is constantly updated with recent photometric data.

The catalogue, now containing only the basic observing data, will also provide a tool for uniform calculation of linear ephemeris in near future. The calculation method has been very well tested and it is based on advanced component analysis (Mikulasek 2007) and robust regression (Mikulasek et al. 2008). It allows combining photometric data acquired in different photometric systems and it eliminates effects of outstanding errors. These ephemerides allow convenient addition of another spectroscopic and spectropolarimetric data, even those with lower accuracy. This will result into the most complex study of the CP stars. Our next goal is to study the stability of the observed periods of stellar variability.

---

[1] In terms of S/N ratio, the precise spectroscopic mean radial velocity measurements, using cross-correlation method (cf. Lehmann et al. 2006, Zverko et al. 2007) in precisely constrained spectral region, is starting to reach the accuracy of classical photoelectric photometry. However, it is very difficult to reach convenient quantitative interpretation of such spectroscopic measurements, so it is mostly impossible to combine them with photometric measurements and use them for diagnostic of secular period changes.

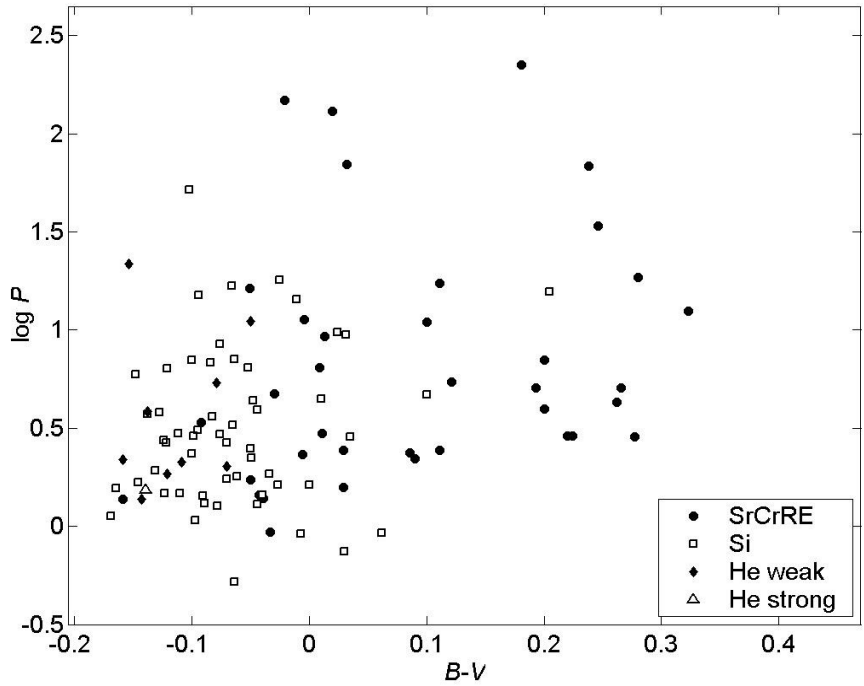

**Figure 2.** Dependency of common logarithm of the period in days on the B-V index with distinguishing various types of mCPod catalogue. It is apparent that cooler mCP stars rotate more slowly.

## 2.2 CP stars with unstable periods and causes of theses changes

Even the most massive MS stars evolve on time scales that are significantly longer then the total time during which we had an opportunity to observe them properly. One would expect that we should not observe any changes of rotational periods. We do not know why CP stars form the complicated structure of regions with different abundances of various chemical elements in their photospheres and how this structure is being maintained. However we guess that those structures last for very long time. Therefore, this should result in an extraordinarily stable period of photometric light changes, as far as those reflect inhomogeneities in the atmospheres of CP stars.

Indeed, more than 90% of CP stars do not indicate any period or light curve changes on longer time scales. Nevertheless, the other cases more or less manifest such peculiar changes and here we try to describe a couple of them.

Recently, Mikulasek et al. (2008) revealed gradual braking of the rotational period $P$ of the young He rich mCP star **V901 Orionis** (HD 37776) with rate of $\dot{P}/P = 4.01(17) \times 10^{-6} \text{yr}^{-1}$. The authors interpret this ongoing period increase as a braking of the star's rotation, at least in its surface layers, due to the momentum loss through events or processes in the extended stellar magnetosphere.

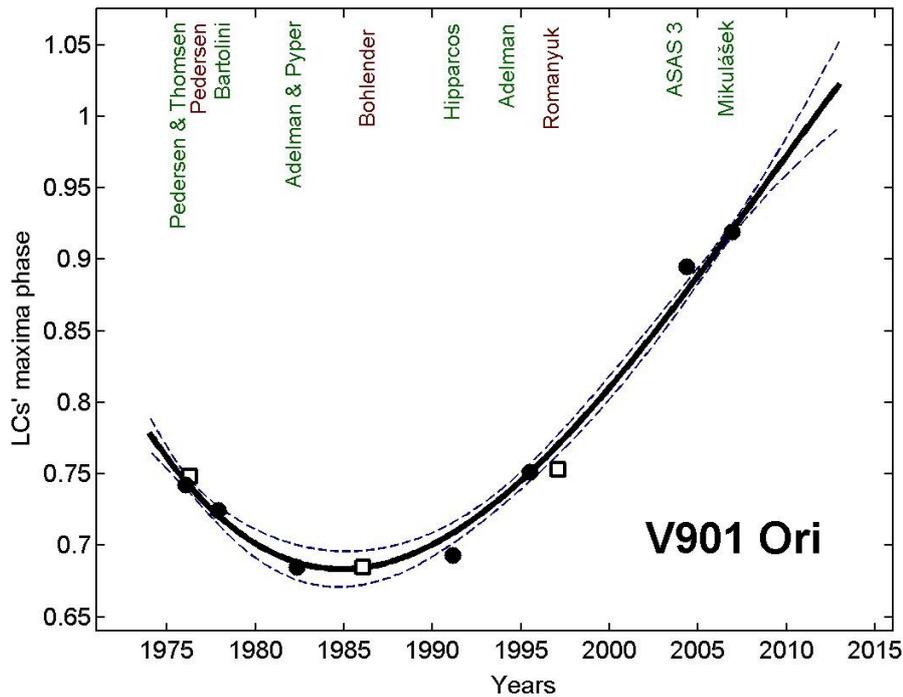

**Figure 3.** The phases of the maxima of the light curves of helium rich mCP star V901 Ori according to linear ephemeris as a function of time undoubtedly document an extraordinary rapid braking of the rotational period of the star. Measurements derived from photometry are denoted by full circles. Open squares correspond to the predicted phases of light maxima derived from the corresponding minima of the equivalent width phase curves of He I lines. The solid curve corresponds to the cubic ephemeris, while the dotted lines indicate the uncertainty of the fit. The figure was taken from Mikulasek et al. (2008).

A gradual increase in the rotation period at a rate of $\dot{P}/P \approx 3\times10^{-7}$ yr$^{-1}$ is exhibited by the rapidly rotating ($P = 0.7279$ d) Si mCP star **SX Arietis** (HD 19832 = 56 Ari) (Musielok 1988, Adelman et al. 2001). For this star the period increase is also accompanied by apparent changes of its *UBV* light curves (Adelman et al. 2001, Ziznovsky et al. 2000). That is why authors expect the observed period changes to be due to free body precession (Shore & Adelman 1976). However, in the case of V901 Ori such an explanation fails as the observed period changes are too large and they are not accompanied by corresponding changes of the shapes of the light curves (Mikulasek et al. 2008).

Two abrupt period increases of one or two seconds have been detected in the very fast rotating ($P = 0.5207$ d) Si mCP star **CU Virginis** (HD 124224) (Pyper et al. 1998, Trigilio et al. 2008) in the years 1985 and 1998. The period jumps have been interpreted as a sudden spinning down of the star as a result of the violent emptying of the inner magnetosphere after the magnetic belt reaches a maximum density (Trigilio et al. 2008). The estimated mean value of $\dot{P}/P$ is $3.7\times 10^{-6}$ yr$^{-1}$, very close to the same value we have derived for V901 Ori. In the case of CU Vir; however, the nature of rotational braking appears different from the gradual spin-down of V901 Ori.

Recently we have treated a mysterious case of the He weak Si CP star **EE Draconis** (HR 7224, HD 177410). Winzer (1974) revealed a medium amplitude photometric variability in EE Dra from 48 *UBV* observations done in 1970-2. Adelman (1997) derived a new period of 1.123095 d based on 616 *uvby* observations taken in seasons 1993/4 and 1994/5 with the

FACPT and *V* of Winzer (1974). Hipparcos obtained 409 *Hp*, $B_T$ and $V_T$ observations in 1989-1992 and introduced the value of the photometric period 1.123248 d.

Then Adelman (2004) reported an unprecedented change in the photometric behaviour of EE Dra. Comparing time series observations taken with FCAPT before 1996 with those taken in 2003 he found that the amplitude of variability increased from typically 0.04 mag to 0.21 mag and that the main period of variability is of order of hundred days. This Adelman's astonishing finding led Lehmann et al. (2006) to start spectroscopic observations. Analysing the double-waved radial velocity curve obtained by the cross-correlation technique on 564 high resolution spectrograms they came up to the rotational period of 1.123248(9) d. They found no further periodicities in the residuals, in particular not in the low frequency region around 100 d period as mentioned by Adelman (2004).

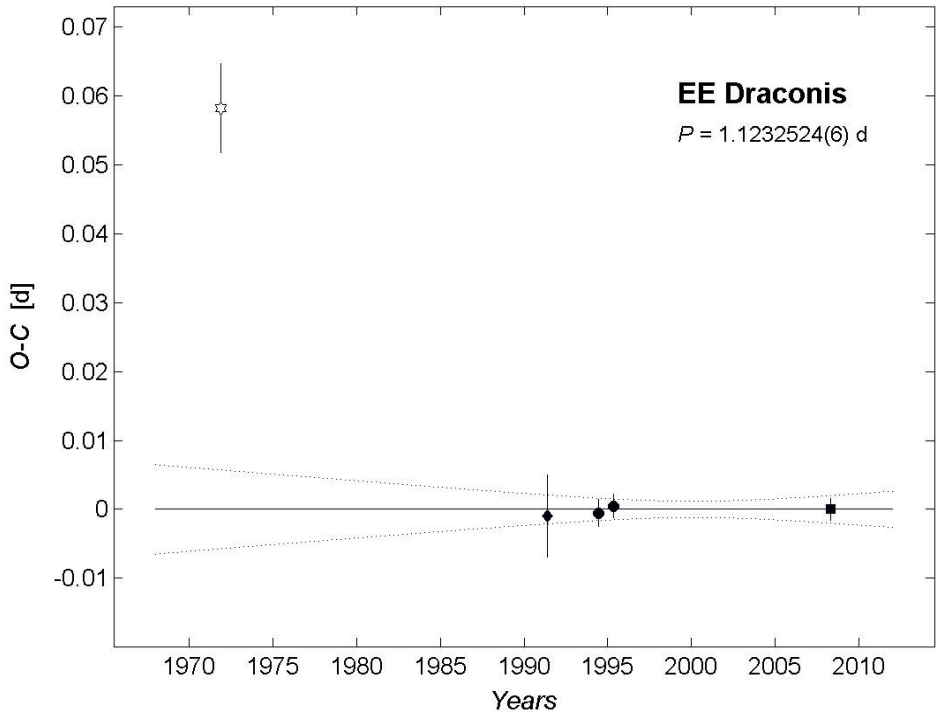

**Figure 4.** O-C diagram of EE Dra plotted according to the new linear ephemeris. Open star denotes observations of Winzer, full diamond corresponds to the Hipparcos measurements, full circles two seasons of Adelman's observations and full square *BV* observations of this paper (Krticka et al. 2008).

In the spring of 2008 we obtained 940 new *BV* measurements and improved the period of variation $P$ = 1.1232524(6) d (Krticka et al. 2008). The found period agrees with those determined by ESA (1998) and Lehmann et al. (2006) and it seems to be constant now. However, we should discuss also the case of the Winzer photometry. The Winzer's light curves seems to be quite correct in their shapes, they are just shifted in respect of our new ephemeris by 0.058(10) d (6 σ). If we accept the shift we could speculate about the possibility of a transient rotational braking which might occur before the Hipparcos mission. Unfortunately, we have no observing evidence for such deceleration between the Winzer's and Hipparchos' epochs. If anyhow we assume a constant rate of the deceleration then we need $\dot{P} \approx 3.4 \times 10^{-9}$ in order to explain the O-C shift of the Winzer's observations. Such a deceleration would manifest itself in the lengthening of the period by 1.8 seconds during the time interval of 17 years (1972-1989). It is not excluded that the strange photometric behaviour of the star in 2003 monitored by Adelman (2004) was a time-retarded violent reaction of upper layers of the star on the change of star's rotational status.

## 3. Conclusions

At present time we know only a few magnetic chemically peculiar stars, which periods and light curves are changing on long-term scale. It is worth to remind that nature of these secular changes, being in progress on time-scales of a couple of thousands rotation periods, differs from star to star. Therefore, it is highly possible that those changes have also different causes. Understanding of these phenomena and their contextualization into the main sequence evolution of hot stars require detailed investigation of the long-term mCP star variability. It is necessary to inspect the largest possible sample of mCP stars, especially those with long-term observing data available from the past. This analysis has to be done in the most complex and accurate way.

Most of the targets are brighter than 7 magnitudes and it is the major bottleneck of this project. Most of the photometric observatories have already switched to CCD detectors and precise photometry of bright stars (in blue colour) is starting to become taboo. The mCP star light variations reach sufficient amplitudes in blue and visual bands; therefore, we would like to encourage all the observers, who are still able to observe bright stars in blue and visual bands with an accuracy several thousandths of magnitude to join this promising project.

**Acknowledgements:** This work was supported by grants GACR 205/06/0217, 205/08/H005, MEB 080832/SK-CZ-0090-07, VEGA 2/6036/27 and by Erasmus programme.